\documentclass[useAMS,usenatbib,a4]{mn2e}
\usepackage{psfig,graphicx,amssymb,amsmath,rotating,nicefrac,longtable,color,natbib}

\def\arcmin{$^{\prime}$}
\def\arcsec{$^{\prime\prime}$}

\def\micron{$\rm{\mu}$m}
\def\microns{$\rm{\mu}$m}
\def\xflux{ergs s$^{-1}$ cm$^{-2}$}
\def\xlum{ergs s$^{-1}$}
\def\psq{$^{-2}$}
\newcommand{\ch}[1]{\textsc{whdfch{\footnotesize#1}}}
\newcommand{\lab}[1]{\textsc{whdf-lab-{\footnotesize#1}}}

\title[Submm observations of X-ray AGN in the WHDF]{Submillimetre observations of X-ray active galactic nuclei in the William Herschel Deep Field}
\author[R. M. Bielby, M. D. Hill, N. Metcalfe and T. Shanks]{R. M. Bielby\thanks{E-mail: rmbielby@gmail.com (RMB)}, M. D. Hill, N. Metcalfe and T. Shanks  \\
Department of Physics, Durham University, South Road, Durham, DH1 3LE, UK
}
\begin{document}
\date{}

\pagerange{\pageref{firstpage}--\pageref{lastpage}} \pubyear{2011}

\maketitle

\label{firstpage}

\begin{abstract}
We investigate the contribution made by active galactic nuclei (AGN) to the high-redshift, luminous, submillimetre (submm) source population using deep ($\leq2$ mJy/beam) Large Apex Bolometer Camera (LABOCA) 870\micron\ observations within the William Herschel Deep Field. This submm data complements previously obtained Chandra X-ray data of the field, from which AGN have been identified with the aid of follow-up optical spectra. From the LABOCA data, we detect 11 submm sources (based on a detection threshold of $3.2\sigma$) with estimated fluxes of $\gtrsim3$ mJy/beam. Of the 11 identified submm sources, we find that 2 coincide with observed AGN and that, based on their hardness ratios, both of these AGN appear to be heavily obscured. We perform a stacking of the submm data around the AGN, which we group by estimated $N_H$ column density, and find that only the obscured ($N_H>10^{22}$ cm\psq) AGN show significant associated submm emission. These observations support the previous findings of Page et al. and Hill \& Shanks that obscured AGN preferentially show submm emission. Hill \& Shanks have argued that, in this case, the contribution to the observed submm emission (and thus the submm background) from AGN heating of the dust in these sources may be higher than previously thought.
\end{abstract}


\section{Introduction}

It is thought that the production of submm emission in the luminous galaxy population observed at $\approx 850$ \micron\ is predominantly driven by star-formation \citep[e.g.][]{2004ApJ...616...71S,2005ApJ...632..736A,2006MNRAS.370.1185P}. However, there is increasing observational evidence for a significant contribution from AGN to the heating of dust, and hence the production of submm emission, in these galaxies. Certainly some fraction of submm sources are observed to emit hard X-rays and there is still the possibility that the observed X-rays are reflected or scattered, which would mean they could have a significantly higher luminosity at 10-30 keV energies than expected.

Obscured AGN (i.e. with $N_H>10^{22}~\mbox{cm}^{-2}$) in particular appear to show high submm fluxes, as identified both from statistical techniques \citep[][]{2011MNRAS.410..762H,2010ApJ...712.1287L} and targeted submm observations of known QSOs \citep[e.g.][]{2008MNRAS.389...45C,2009ApJ...706..184M}. \citet{2004ApJ...611L..85P} presented samples of obscured and unobscured QSOs and suggested that there may be a difference between these two classes in the submm, with the obscured sources showing higher submm fluxes (see also \citealt{2005MNRAS.356.1571M} for a fainter example of Type 2 QSO). This is a crucial observation since it implies that the unified AGN model -- where X-ray obscured QSOs are interpreted as being viewed at a different angle than unobscured QSOs -- may be incomplete. Instead, X-ray obscured QSOs may represent an earlier stage in QSO evolution, where the black-hole formed within a dusty starburst galaxy, as suggested by \citet{2004ApJ...611L..85P}. This would account for an instrinsic difference between the submm properties of obscured and unobscured AGN.

Recently, \citet{2011MNRAS.413.2791C} presented an analysis of a $z=1.82$ QSO associated with submm emission and which is surrounded by an over-density of submm galaxies (both QSOs and submm galaxies are strong tracers of large scale structure, e.g. \citealt{1990ApJ...348...38S,1991ApJ...371...49E,2009MNRAS.393.1573A,2009ApJ...691..560C,2010A&A...523A..66B,2010ApJ...722..102S,2011MNRAS.416.2041M}). Based on spectral energy distribution (SED) fitting, they conclude that the emission in the mid-infrared (MIR) range is dominated by reprocessed AGN emission, whilst the submm is dominated by a starburst contribution. A similar result was found for a $z\approx4$ QSO by \citet{2009MNRAS.395.1905C}. However, \citet{2011MNRAS.414.1875H} fit galaxy counts and colours using
optically defined Pure Luminosity Evolution (PLE) models, where dust
reradiates absorbed optical light into infrared spectra composed of
local galaxy templates, across the MIR bands from 3.6\microns\ to 8\microns\ and up to a redshift of at least $z=2.5$. They show that a significant contribution from AGN to the $250-870~$\microns\ SED would remove the need to invoke a top-heavy IMF for high-redshift starburst galaxies. It may therefore be that reprocessed emission from AGN could be more important than previously thought.

Interestingly, a significant AGN contribution to the submm population could have relevance for the origin of the hard X-ray (i.e. 10-30 keV) background. At softer energies (i.e. $\lesssim10$ keV), much of the X-ray background (XRB) is resolved into sources,  usually  AGN \citep{1991Natur.353..315S}. At harder energies therefore the expectation is that heavily obscured AGN may form the background \citep[e.g.][]{1995A&A...296....1C,2007A&A...463...79G}. However, the hard X-ray sources that have so far been identified by Chandra and XMM are skewed towards  lower redshift, peaking at $z<1$ and hence have somewhat lower luminosities, $L_X\approx10^{42}-10^{44}$ erg/s \citep[e.g][]{2003AJ....126..539A}. Most studies conclude that the contributing sources so far detected only comprise $<25$\% of the XRB \citep{2005MNRAS.357.1281W,2005ApJ...625...89K,2006MNRAS.368.1735W,2007ApJ...670..173D,2007ApJ...661L.117H}. Should submm sources contain obscured AGN, this may go some way to explaining the missing XRB flux.

There is some evidence that different populations of high redshift galaxies host obscured AGN. For example, \citet{2007ApJ...670..173D} reported that a number of $BzK$ selected galaxies show a thermal, warm dust excess at 24 microns resulting from faint, hard X-ray sources. The suggestion is that these are obscured QSOs, but again these X-ray sources may be too faint to account for any more than 25\% of the background at 10-30 keV (whilst we note that \citealt{2011ApJ...738...44A} have recently reanalyzed the \citealt{2007ApJ...670..173D} galaxy sample with deeper X-ray data and found a factor of $\gtrsim10$ fewer obscured AGN than originally suggested by \citealt{2007ApJ...670..173D}). In terms of submm galaxies, stacked X-ray spectra show a broad Fe $K_\alpha$ line \citep{2005ApJ...632..736A}, whilst many of the optical spectra of submm galaxies show evidence for broad lines. Additionally, high ionisation lines indicative of AGN activity have also been detected at FIR wavelengths in Herschel SPIRE FTS observations of a $z\approx3$ submm galaxy \citep{2011MNRAS.415.3473V}. Finally, \citet{2001MNRAS.323...67B} suggested that the obscured QSOs might explain the bright, $\approx$5mJy, submm number counts if the QSOs had a temperature of $\approx$30K \citep[see also][]{1999MNRAS.305L..59A} and this is at least consistent with the temperatures being reported for the SCUBA Half-Degree Extragalactic Survey (SHADES) submm sources by \citet{2008MNRAS.389...45C}. Thus, it may be that submm galaxies account for a further fraction of the X-ray background at 10-30 keV.

We have undertaken a survey designed to measure 870\micron\ fluxes for a sample of known quasars with the aim of comparing the properties of obscured versus unobscured AGN. This has been performed in the William Herschel Deep Field \citep[WHDF;][]{metcalfe95,metcalfe01,metcalfe06}, which is especially suitable for this survey thanks to a significant sample of spectroscopically confirmed quasars within a small, easily observable area. This includes both unobscured and heavily obscured sources matched in redshift and luminosity \citep{2004PhDT.VallbeMumbru}. Although not capable of producing definitive results on their own, the observations we present here are a crucial step towards identifying the contribution of AGN to the production of the submm emission in submm sources and thus, this population's contribution to the submm and X-ray backgrounds. This paper therefore presents the results from the submm survey, focussing on statistical analyses that complement the analyses in \citet{2011MNRAS.410..762H}, and which will form the basis for further multi-wavelength treatment of this issue.

\section{The William Herschel Deep Field} \label{s-whdf}

The William Herschel Deep Field is a $\approx7$\arcmin$\times7$\arcmin\ area centred at $\approx$ $00^h20^m$ $+00^{\circ}$ (J2000) which has a wealth of multiwavelength data and has been extensively studied over the past 15 years \citep{metcalfe95,2000MNRAS.311..707M,2000MNRAS.318..913M,metcalfe01,metcalfe06}.

%\begin{figure}
%\centering
%\includegraphics[width=8cm]{figs/whdf.eps}
%\caption[A true colour image of the William Herschel Deep Field]{A true colour image of the William Herschel Deep Field from the $U$, $B$ and $R$ band frames. Image credit: N. Metcalfe.}
%\label{f-whdfcol}
%\end{figure}

The field has ultradeep, ground-based optical $UBRIZ$ imaging from the William Herschel Telescope in La Palma (reaching $B<27.9$) as well as near-infrared $H$ and $K$ imaging from Calar Alto and the UK Infrared Telescope and deep, high-resolution $I$-band imaging from the \emph{Hubble Space Telescope}'s Advanced Camera for Surveys.

In addition to this comprehensive optical/NIR coverage, the WHDF also has \emph{Chandra} X-ray coverage, reaching a depth of $\approx10^{-15}$ \xflux\ over the whole area with a total integration time of $\approx70$ ksec. These observations were undertaken between November 2000 and January 2001. 170 X-ray sources were detected at $\ge2\sigma$ significance, of which 69 were at $\ge3\sigma$ and 36 at $\ge5\sigma$. Spectroscopic follow-up of some of these X-ray sources was subsequently performed and these observations are described by \citet{2004PhDT.VallbeMumbru} and in \S\ref{ss-qsosamples} below. Further spectroscopic data (of star-forming and passive galaxies respectively) has also been presented by \citet{2007ApJ...668..846B} and \citet{2009MNRAS.393.1467F}.

The most recent additions to the WHDF data are (\emph{a}) an 870\micron\
submm survey, which we undertook between 2008 and 2009 and which is
described in this paper, and (\emph{b}) radio observations at 8.4 GHz
($\lambda=3.6$cm) acquired with the Expanded Very Large Array (EVLA) in
New Mexico in 2010. The reduction of these radio data is under way.

%The EVLA survey will complement very well the submm observations. The
%well established FIR-radio correlation \citep{1992ARA&A..30..575C} means
%that radio data are frequently employed to determine more precise
%positions for submm sources \citep[e.g.][]{2002MNRAS.337....1I}, since
%the beamsize of submm instruments is very large ($\sim20$\arcsec). It
%is assumed that where a radio source lies within the error box of a
%submm source position, it can reliably be taken to be the counterpart
%of the submm source. Radio interferometers like the EVLA are able to
%determine source positions with far greater precision than the
%$\pm10$\arcsec\ accuracy of the submm data. To this end, therefore, we
%have acquired 35 hours of EVLA X-band observations, expected to reach an
%rms noise level of $\approx2$ $\mu$Jy, sufficient to detect our submm
%sources. The reduction of these data is underway.

\section{submm observations and data reduction}

\subsection{Observations}

We have acquired 21 hours of observations of the WHDF with the Large Apex Bolometer Camera \citep[LABOCA;][]{2009A&A...497..945S} on the 12 m APEX telescope \citep{2006A&A...454L..13G}. The LABOCA instrument  comprises 295 semiconducting composite bolometers arranged in a series of concentric hexagons. LABOCA is sensitive to radiation in a passband centred at 870\microns, with a FWHM of $\approx150$\microns.

\begin{figure*}
\centering
\includegraphics[width=0.8\textwidth]{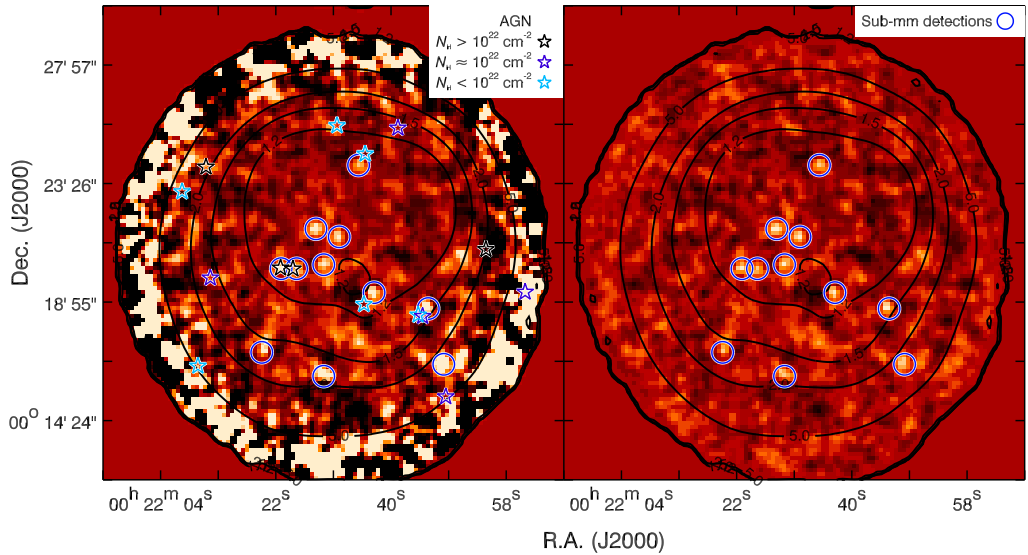}
\caption[LABOCA 870\micron\ flux and SNR maps of the WHDF]{\emph{(a)} LABOCA 870\micron\ intensity map of the WHDF, with 11 sources detected at $\ge3.2\sigma$ marked by open blue circles. The overlayed contours show the redcued noise map, with contour levels marked in mJy/beam. Spectroscopically identified QSOs (listed in Table~\ref{t-qsos}) are also plotted, coded by their estimated hydrogen column density. The reduced off-axis sensitivity of LABOCA produces the increase in noise at the edges of the frame. \emph{(b)} LABOCA signal-to-noise map of the WHDF, with the same 11 sources marked. Again the contours show the reduced noise map.}
\label{f-whdfmaps}
\end{figure*}

The LABOCA beam has a FWHM of 18.6\arcsec and the total field of view (FoV) of the detector is 11.4\arcmin. The WHDF covers a region of $\approx7$\arcmin$\times7$\arcmin, as noted above, so this field  fits well into the FoV of LABOCA. The LABOCA detectors do not form a contiguous array, however, so to achieve full sampling of the field we carried out our observations using a standard spiral raster map pattern. In this process, the centre of the array is moved in a spiral pattern and whilst being shifted laterally in a raster configuration, in order to fully sample the FoV. 

Our observations were carried out in two separate observing runs\footnote{ESO programme IDs 081.A-0897(A) and 083.A-0707(A)}, the first on 29-30th August 2008 and the second on 11-12th May 2009. The observing conditions were very good, with the range of zenith opacities, $\tau_z$, spanning 0.17 to 0.23 in the first run and 0.17 to 0.28 in the second (these being calculated as standard from a linear combination of LABOCA 870 \micron\ skydip results and opacities determined from the APEX radiometer).

\subsection{Data reduction}

Initial data reduction was performed using the standard \textsc{boa} pipeline software (\citealt{2010BoA}\footnote{www.mpifr-bonn.mpg.de/div/submmtech/software/boa/boaman.pdf}) using the recommended sequence for `weak sources'. Counts from the detector are first converted into an output voltage, which can then be converted into a flux density, in standard units of Jy beam$^{-1}$, via a voltage-flux relation determined empirically during LABOCA's commissioning. Sky removal was performed in BoA using the iterative {\it medianNoiseRemoval} command, which computes and corrects for the relative gains for a channel with respect to the mean signal. In addition to the usual de-spiking, data flagging, correlated noise removal and Fourier space filtering during the reduction, a DC offset (modelled by a first order polynomial) is removed from each scan for each bolometer. A further correction is then applied to account for the zenith opacity at the time of observing. The zenith opacity, $\tau_z$, is a measure of how much incoming radiation is absorbed by the atmosphere and is determined by `skydip' observations (calibration exposures of the sky). Flux calibration was performed using observations of primary and secondary calibrators, with Uranus and Neptune being used as primary calibration sources during our observing run. The result of this process is a series of exposure maps for each scan of the target field. The maps were then co-added using BoA, which weights the signal from each scan image by $1/\sigma^2$, where $\sigma$ is the pixel-noise and is estimated by adding in quadrature the noise levels for each bolometer that contributes to a given pixel. The final output is a combined intensity (flux) map, as well as corresponding rms noise and signal-to-noise (SNR) maps. Each of the maps has a pixel scale of $9''.1$/pixel.

This data reduction process was carried out separately for the data from each of the two observing runs. As a final stage in the reduction process, therefore, we combined the maps, weighting them by the noise. The final intensity map was produced according to Eqn.\ \ref{eq-intcomb} and the final noise map according to Eqn.\ \ref{eq-noisecomb}.

\begin{equation}
\label{eq-intcomb}
%I=\left[2\times\middle(\frac{I_1}{\sigma_1^2}+\frac{I_2}{\sigma_2^2}\middle)\middle] \middle/ \middle[\frac{1}{\sigma_1^2}+\frac{1}{\sigma_1^2}\right]
I=\left(\frac{I_1}{\sigma_1^2}+\frac{I_2}{\sigma_2^2}\middle) \middle/ \middle(\frac{1}{\sigma_1^2}+\frac{1}{\sigma_1^2}\right)
\end{equation}

\begin{equation}
\label{eq-noisecomb}
\sigma=1\left/ \sqrt{\frac{1}{\sigma_1^2}+\frac{1}{\sigma_2^2}}\right.
\end{equation}

\noindent where $I$ is intensity, $\sigma$ is rms noise and subscripts 1 and 2 indicate the first and second observing runs. A signal-to-noise map was produced by taking the ratio of the two, and each of the maps was then Gaussian-smoothed using the $18''.6$ beam profile, which gave a final resolution for the smoothed maps of $27''$. The smoothing was comparable to that done by \citet{2009ApJ...707.1201W}, who have shown that flux estimates from similarly smoothed maps are consistent with non-smoothed fluxes.

\section{LABOCA data}

The final intensity and SNR maps are shown in Fig.\ \ref{f-whdfmaps}. It is clear from Fig.\ \ref{f-whdfmaps} that the noise increases substantially at large off-axis angles; in the SNR map (right-hand panel), this increase in the noise manifests itself as a dearth of sources around the edges of the frame. In the central 100 arcmin$^2$ area, the maps are very similar, with points of bright intensity having corresponding peaks in the SNR map, an indication that the noise level is relatively uniform across the field centre. 

\subsection{Noise level}

Across the central 100 arcmin$^2$ region of the field the data reach an rms noise level of $\lesssim2$ mJy beam$^{-1}$, making the WHDF one of the deepest submm fields observed to date (\emph{cf}. figure 6 of \citealt{2009ApJ...707.1201W}). 

To check the reliability of the pipeline-reduced noise map we have produced a `standard deviation map', by measuring the standard deviation of the intensity map in a series of annuli from the field centre, having masked out the bright sources (see Section~\ref{sec:labsources}). If the pipeline has worked successfully, this standard deviation map should be comparable to the BoA-produced noise map. 

\begin{figure}
\centering
\includegraphics[width=\columnwidth]{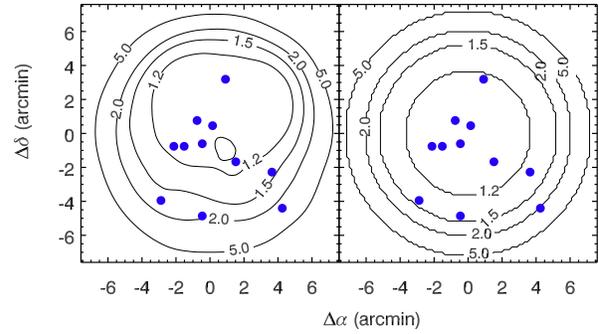}
\caption[LABOCA 870\micron\ noise contour maps of the WHDF]{ Noise contour maps of the WHDF, with the rms noise level marked at 1.2, 1.5, 2.0 and 5.0 mJy beam$^{-1}$, using \emph{(a)} the pipeline-reduced noise map and \emph{(b)} the standard deviation of the background flux measured in radial annuli. The image pixel size is $9.''1$. $3.2\sigma$ LABOCA sources are marked as blue circles. The maps generally agree well (see also Fig.\ \ref{f-noiseprof}). Most of our LABOCA sources are detected within the central region where the noise is lowest.}
\label{f-noise}
\end{figure}

\begin{figure}
\centering
\includegraphics[width=\columnwidth]{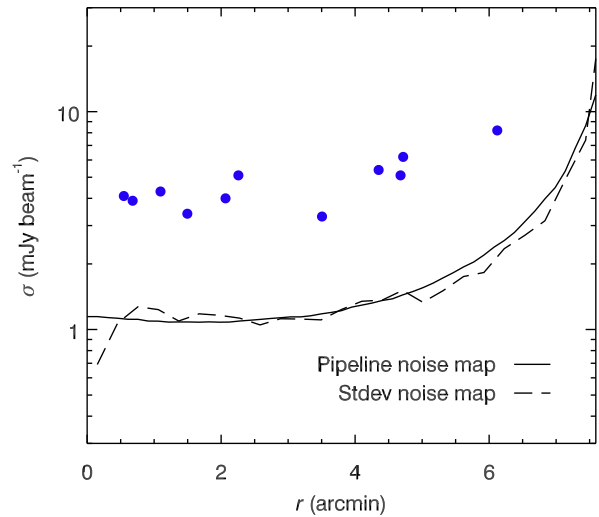}
\caption[Radial noise profile of the WHDF LABOCA map]{Radial noise profile of the WHDF showing the rms noise level, as a function of radius from the centre, for both the pipeline-reduced and standard deviation noise maps, described in the main text. $3.2\sigma$ LABOCA sources are marked (blue circles) at their respective source flux and radial distance from the centre of the field.}
\label{f-noiseprof}
\end{figure}

In Figs.\ \ref{f-noise} and \ref{f-noiseprof}, we compare the contours and radial profiles, respectively, of the two maps. For the pipeline-reduced map, the noise profile shown in Fig.\ \ref{f-noiseprof} is the median of 120 profiles measured radially at $3'$ intervals. The agreement between the two noise maps is very good --- this is particularly clear from Fig.\ \ref{f-noiseprof} --- suggesting that the noise has been reliably estimated for our field and that our quoted depth of 2 mJy beam$^{-1}$ over an area of 100 arcmin$^2$ is robust.

\subsection{LABOCA sources}
\label{sec:labsources}
\begin{figure}
\centering
\includegraphics[width=\columnwidth]{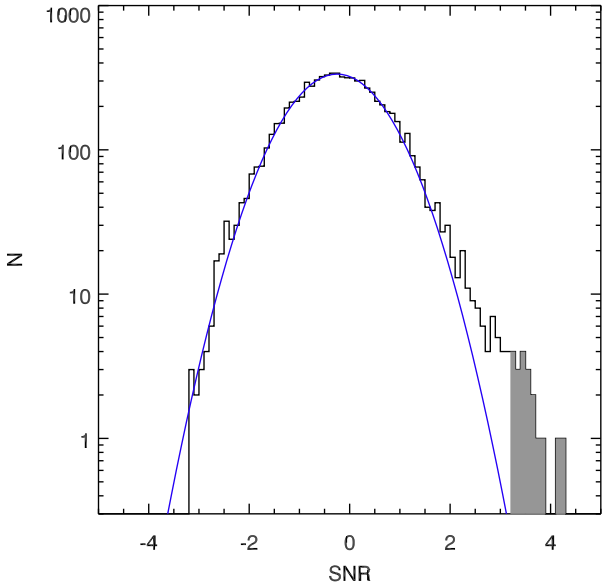}
\caption[SNR histogram for the WHDF LABOCA data]{A histogram of pixel values from the SNR map. A Gaussian has been empirically fit to the profile --- most of the pixels lie within this distribution, however a significant positive excess appears due to the presence of sources. The shaded region indicates those pixels lying above the $3.2\sigma$ threshold we have set for source detection.}
\label{f-snrhist}
\end{figure}

The field contains 11 significant submm sources, circled in Fig.\ \ref{f-whdfmaps}. We select the sources from the SNR map, with a criterion of $S/N\ge3.2$. This significance was chosen by comparing the map to its inverse: there are no negative spikes in the SNR map with a magnitude of $3.2\sigma$. This is illustrated in Fig.\ \ref{f-snrhist}, which shows a pixel value histogram for the SNR map, with a Gaussian fit (blue curve). The map shows a strong positive excess while the negative side of the distribution follows the Gaussian curve well. Based on the Gaussian fit, we estimate that the data would contain $\approx0.2$ spurious peaks above our chosen SNR limit within the central 100 arcmin$^2$ region.

The positions, SNRs and fluxes of the 11 sources are summarised in Table~\ref{t-laboca}. We note that the quoted fluxes have not been corrected for the flux bias effect \citep{2008MNRAS.384.1597C,2009MNRAS.393.1573A}, however we do not consider this necessary for the scope of this paper. The positions of the 11 sources are indicated on Figs.~\ref{f-whdfmaps} and \ref{f-noise} and their fluxes are shown in comparison to the background noise in Fig.\ \ref{f-noiseprof}. Most of the sources lie within the $<2$ mJy/beam central area of the field, with 7/11 located within the $\sigma\le1.2$ mJy contour. 

%The source densities are $N(S) = 653\pm197$ deg$^{-2}$, $348\pm123$ deg$^{-2}$ and $174\pm78$ deg$^{-2}$ at limiting fluxes of $S = 3.2$ mJy, $4.0$ mJy and $5.0$ mJy respectively. The faintest point lies just above the LESS $N(>S)$ in Fig. 12 of \citet{2011MNRAS.410..762H}.

\begin{table*}
\centering
\caption[WHDF LABOCA 870 \micron\ sources detected at $\ge3.2\sigma$]{WHDF LABOCA 870 \micron\ sources detected at $\ge3.2\sigma$.}
\begin{tabular}{cccccc}
\\
\hline
ID & RA & Dec & SNR & $S_{870}~^a$ & $\Delta\theta$\\
 & \multicolumn{2}{c}{(J2000)} &  & (mJy/beam) & ($''$)\\
\hline
WHDF-LAB-01 & 00:22:37.55 & +00:19:16.8 & 4.3 & 5.1 & 9.4\\
WHDF-LAB-02 & 00:22:28.44 & +00:21:42.6 & 4.2 & 4.3 & 9.6\\
WHDF-LAB-03 & 00:22:46.06 & +00:18:40.3 & 3.8 & 5.4 & 10.7\\
WHDF-LAB-04 & 00:22:29.66 & +00:20:20.5 & 3.7 & 4.1 & 10.9\\
WHDF-LAB-05 & 00:22:22.97 & +00:20:11.4 & 3.6 & 4.0 & 11.3\\ % whdfch008
WHDF-LAB-06 & 00:22:32.09 & +00:21:24.3 & 3.6 & 3.9 & 11.3\\
WHDF-LAB-07 & 00:22:48.49 & +00:16:32.8 & 3.5 & 8.2 & 11.6\\
WHDF-LAB-08 & 00:22:29.66 & +00:16:05.4 & 3.4 & 6.2 & 11.9\\
WHDF-LAB-09 & 00:22:19.90 & +00:17:00.1 & 3.2 & 5.1 & 12.7 \\
WHDF-LAB-10 & 00:22:35.16 & +00:24:08.3 & 3.2 & 3.3 & 12.7\\
WHDF-LAB-11 & 00:22:25.40 & +00:20:11.4 & 3.2 & 3.4 & 12.7\\ %whdfch007
\hline
\multicolumn{6}{l}{$^a$ we note that listed fluxes are measured fluxes and not corrected for flux bias.} \\
\end{tabular}
\label{t-laboca}
\end{table*}

\subsection{Completeness}

\begin{figure}
\centering
\includegraphics[width=\columnwidth]{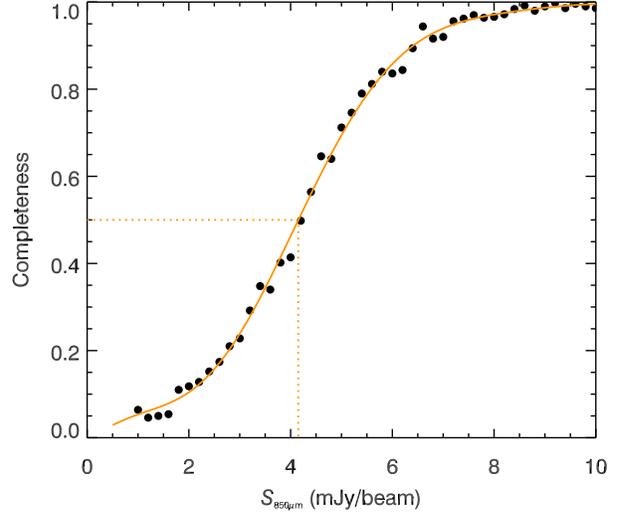}
\caption[submm completeness]{The completeness of the submm data ($<2$ mJy/beam region only) estimated based on simulated sources placed in the submm image. The solid line shows a polynomial fit to the completeness estimates as a function of source brightness, with 50\% completeness marked by the dotted line at $S_{870\mu m}=4.2$ mJy/beam.}
\label{fig:cmpltnss}
\end{figure}

We estimate the completeness of our submm observations using simulated sources placed in the final submm map across a range of fluxes. At each flux level, 500 sources are individually added to the map in turn and the detection analysis re-performed for each source. This is limited to the $<2$ mJy/beam region only and excluding regions within 2 pixels of any of the 11 detected sources. The result is shown by the filled black circles in Fig.~\ref{fig:cmpltnss}, with the curve showing a polynomial fit to the result. Based on this analysis, we find the data to be 50\% complete at the 4.3 mJy/beam level (shown by the dotted line). 

\section{submm observations of WHDF quasars}

\subsection{Obscured and unobscured quasar samples} \label{ss-qsosamples}

In \S\ref{s-whdf} we briefly summarised the \emph{Chandra} observations of the WHDF, which yielded detections of 170 X-ray sources. In 2001 and 2002, 36 of these WHDF X-ray sources were targetted in 10 hours of spectroscopic follow-up with the LDSS2 multi-object spectrograph on the 6.5m Magellan-1 (Walter Baade) telescope at the Las Campanas Observatory in Chile. These were optical observations, using a grism centred at $\lambda\approx5500$\AA. 

A stated aim of the LDSS2 observations was to investigate non-quasar X-ray sources, so the principal targets for spectroscopic follow-up were X-ray sources whose optical counterparts did not appear especially point-like \citep{2004PhDT.VallbeMumbru}. Nevertheless, 15 of the targets were spectroscopically confirmed as QSOs, and a further 2 were classified as being either QSOs or narrow emission line galaxies (NELGs). In this section we take the 15 confirmed QSOs as our sample of WHDF quasars; details of these sources are given in Table \ref{t-qsos}.

\begin{table*}
\centering
\caption[Spectroscopically confirmed quasars in the WHDF]{Spectroscopically confirmed quasars in the WHDF. The positions, 0.5--10 keV fluxes (in \xflux), X-ray hardness ratios and spectroscopic redshifts are given. obscured quasars are indicated in bold.}
\label{t-qsos}
\begin{tabular}{cccccc}
\\
\hline
ID$^a$ & R.A. & Dec. & $S_{0.5-10}$ & $HR$ & $z$ \\
       & \multicolumn{2}{c}{(J2000)} & (erg/cm$^2$/s)    &   &   \\
\hline
WHDFCH005 & 00:22:35.963 & 00:18:50.04  & $5.62\times10^{-14}$  & $-0.60 \pm 0.08$  & $0.52$ \\% & QSO  \\
%WHDFCH006 & 00:22:32.584 & 00:23:27.84  & $9.10\times10^{-16}$  & $-0.88 \pm  0.15$  & $0.38$ & NELG/QSO \\
\textbf{WHDFCH007} & 00:22:24.821 & 00:20:10.94  & $1.17\times10^{-14}$  & $+0.82 \pm 0.33$  & $1.33$  \\% & QSO \\
\textbf{WHDFCH008} & 00:22:22.884 & 00:20:13.24  & $3.62\times10^{-15}$  & $-0.20 \pm  0.24$  & $2.12$ \\% & QSO \\
WHDFCH016 & 00:22:45.164 & 00:18:22.64  & $1.44\times10^{-14}$  & $-0.43 \pm 0.14$  & $1.73 $  \\% & QSO \\
WHDFCH017 & 00:22:44.468 & 00:18:25.64  & $3.22\times10^{-13}$  & $-0.55 \pm  0.03$  & $0.40 $ \\% & QSO \\
WHDFCH020 & 00:22:36.142 & 00:24:33.84  & $1.09\times10^{-14}$  & $-0.55 \pm 0.17$  & $0.95 $  \\% & QSO \\
WHDFCH036 & 00:22:31.734 & 00:25:38.84  & $6.26\times10^{-14}$  & $-0.48 \pm 0.06$  & $ 0.83 $ \\% & QSO \\
\textbf{WHDFCH044} & 00:22:55.092 & 00:20:55.74  & $2.66\times10^{-14}$  & $+0.60 \pm 0.14$  & $0.79 $  \\% & QSO \\
WHDFCH048 & 00:22:41.297 & 00:25:33.34  & $2.15\times10^{-14}$  & $-0.43 \pm 0.10$  & $1.52 $   \\% & QSO \\
WHDFCH055 & 00:22:11.862 & 00:19:50.44  & $2.17\times10^{-14}$  & $-0.23 \pm 0.06$  & $0.74 $ \\% & QSO \\
%WHDFCH060 & 00:22:28.453 & 00:15:01.94  & $1.93\times10^{-14}$  & $-0.33 \pm 0.10$  & $0.36 $  & NELG/QSO \\
WHDFCH090 & 00:22:48.795 & 00:15:18.74  & $4.83\times10^{-14}$  & $-0.41 \pm 0.06$  & $1.32 $  \\% & QSO \\
\textbf{WHDFCH099} & 00:22:11.187 & 00:24:04.13  & $8.84\times10^{-15}$  & $+0.11 \pm 0.18$  & $0.82 $  \\%  & QSO \\
WHDFCH109 & 00:22:09.917 & 00:16:28.94  & $6.69\times10^{-14}$  & $-0.55 \pm 0.08$  & $0.57 $  \\%  & QSO \\
WHDFCH110 & 00:22:07.433 & 00:23:07.74  & $2.20\times10^{-14}$  & $-0.43 \pm 0.14$  & $0.82 $ \\%   & QSO \\
WHDFCH113 & 00:23:01.247 & 00:19:18.01  & $5.99\times10^{-15}$  & $-0.44 \pm 0.45$  & $2.55 $ \\%   & QSO \\
\hline
\multicolumn{6}{l}{$^a$ as in \citet{2004PhDT.VallbeMumbru}} \\
\\
\end{tabular}
\end{table*}

\begin{figure}
\centering
\includegraphics[width=\columnwidth]{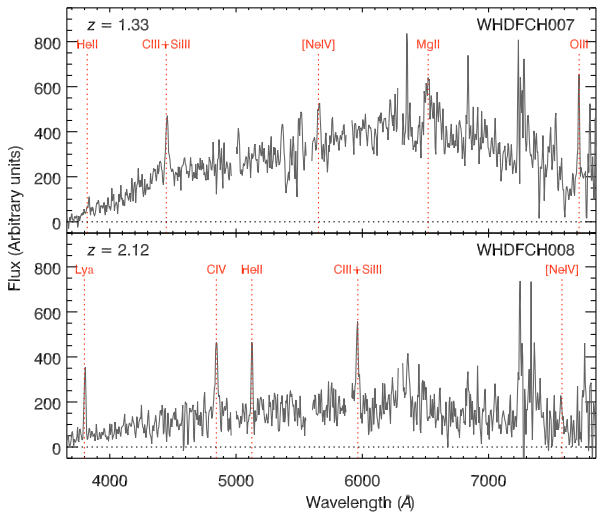}
\caption[]{Optical spectra of \ch{007} (top) and \ch{008} (bottom) acquired with LDSS2 on the Magellan telescope. Both objects are associated with submm sources in our sample. Broad MgII emission is observed in the spectrum of \ch{007}. Only narrow emission lines are detected for \ch{008}, which was classified as a type 2 QSO by\citet{2004PhDT.VallbeMumbru}.}
\label{fig:magellanspec}
\end{figure}

The QSO classification was made by \citet{2004PhDT.VallbeMumbru} on the basis of both X-ray luminosity ($L_{\mathrm{X}}>10^{44}$ \xlum) and optical emission lines. Most of the sources classed as QSOs showed broad emission lines, including Ne\textsc{v} $\lambda$1240, Si\textsc{iv} $\lambda$1400, N\textsc{iv} $\lambda$1486, He\textsc{ii} $\lambda$1640, O\textsc{iii}] $\lambda$1663 and/or N\textsc{iii} $\lambda$1750, all of which are AGN indicators. If broad lines were detected the source was classified as a type 1 quasar, for example \ch{007} which is shown in the top panel of Fig.\ \ref{fig:magellanspec}. One source was classified as a type 2 quasar on the basis of strongly detected narrow emission lines (\ch{008}); its spectrum is shown in the bottom panel of Fig.\ \ref{fig:magellanspec}.

\begin{figure}
\centering
\includegraphics[width=\columnwidth]{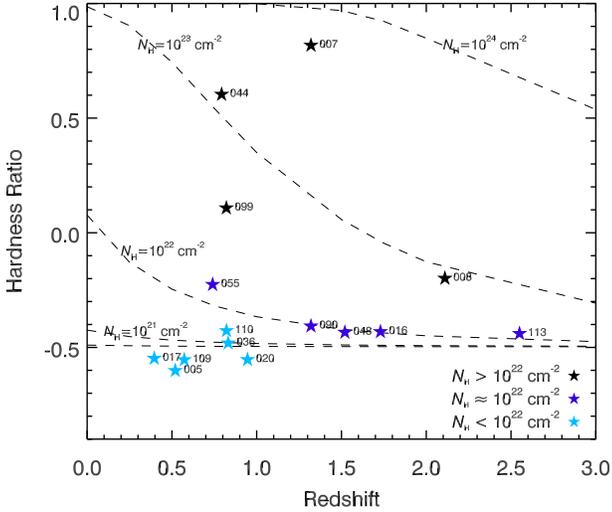}
\caption[Hardness ratio versus redshift for WHDF QSOs]{Hardness ratio versus redshift. The curves show predicted tracks for obscured QSOs at different column densities, which are indicated on the figure (in units of cm\psq). An intrinsic $\Gamma=2$ power-law spectrum is assumed. Below $N_{\mathrm{H}}=10^{21}$ cm\psq\ the lines become indistinguishable. WHDF X-ray QSOs are marked; each is labelled with its ID (with the \ch{} prefix omitted). On the basis of this figure we classify the QSOs into 3 categories: heavily obscured ($N_{\mathrm{H}}>10^{22}$ cm\psq; black), mildly obscured ($N_{\mathrm{H}}\approx10^{22}$ cm\psq; blue) and unobscured ($N_{\mathrm{H}}<10^{22}$ cm\psq; cyan).}
\label{f-hr-z}
\end{figure}

The level of obscuration of an AGN can be estimated directly from the X-ray spectrum using the hardness ratio, $HR$. This is defined as $HR = (H-S)/(H+S)$, where $H$ and $S$ represent the photon counts in the hard (2 - 8 keV) and soft (0.5 - 2 keV) X-ray bands respectively. Increasing levels of absorbing hydrogen column density produce greater absorption in the soft band than in the hard, so leading to larger values of $HR$. Assuming a given intrinsic spectral slope, $\Gamma$, the absorbing hydrogen column density can be estimated. Fig.\ \ref{f-hr-z} shows hardness ratio against redshift for our quasar sample, compared to the predicted tracks for quasars at different absorbing column densities (an intrinsic photon index of $\Gamma=2$ is assumed).

The locus of the quasars is at $HR\approx-0.5$, consistent with the model that predicts $HR=-0.5$ for essentially all unobscured QSOs. Only four of the 15 QSOs are harder than $HR=-0.2$; these 4 sources --- \ch{007}, -{\footnotesize008}, -{\footnotesize044} and -{\footnotesize099} --- are expected to be highly obscured. \ch{007} and \ch{044} have extremely hard spectra with $HR\gtrsim0.6$, corresponding to an apparent photon index of $\Gamma\approx-1$. Such sources are relatively rare, with for example only 2 $HR\approx0.6$ objects being reported in the $\approx2$ deg$^2$ \emph{XMM-Newton} COSMOS survey \citep{2007ApJS..172..368M} (which we note is less deep by a factor of $\approx2-3\times$ than the WHDF X-ray data used here, whilst they do not present $HR$ values with errors of $>0.3$).

Based on Fig.\ \ref{f-hr-z} we divide the WHDF quasar sample into three groups: heavily obscured ($N_{\mathrm{H}}>10^{22}$ cm\psq), mildly obscured ($N_{\mathrm{H}}\approx10^{22}$ cm\psq) and unobscured ($N_{\mathrm{H}}<10^{22}$ cm\psq); these are marked in both Fig.~\ref{f-whdfmaps} and Fig.~\ref{f-hr-z} by black, blue and cyan stars respectively.

\subsection{submm properties}

\subsubsection{Possible counterparts}

We now look for associations between the submm sources found in the LABOCA data and the spectroscopically confirmed QSOs described above. For the purposes of this analysis, we restrict the sample to only those QSOs that lie within the $2$ mJy/beam contours of the pipeline reduced submm noise map shown in the left hand panel of Fig.~\ref{f-noise}. This leaves two $N_H>10^{22}\mbox{ cm}^{-2}$, three $N_H\approx10^{22}\mbox{ cm}^{-2}$ and four $N_H<10^{22}\mbox{ cm}^{-2}$ QSOs.

In order to identify coincident sources, we use a maximum separation between a QSO and a submm source of $2.5\Delta\theta$, where $\Delta\theta$ is the positional uncertainty on the given submm source (listed in Table~\ref{t-laboca}). Given the signal-to-noise limit on our submm catalogue of $S/N=3.2\sigma$, this gives a maximum possible separation of $2.5(0.6\theta(S/N)^{-1})=12.7''$ \citep{2007MNRAS.380..199I}, with the smoothed map resolution of $\theta=27''$. The majority of the submm sources will have a separation constraint somewhat smaller than this however, given their higher signal-to-noise.

In order to evaluate the significance of a given alignment, we use the corrected Poisson probability, $P$, as employed by \citet{1986MNRAS.218...31D}. As such, we estimate the probability of a chance alignment using the observed integrated sky density of X-ray sources, $N(>S)$, as a function of the soft X-ray flux, $S_{0.5-2}$, which we calculate based on the power-law fit given by \citet{2004PhDT.VallbeMumbru} for the WHDF sources:

\begin{equation}
	\log{N(>S_{0.5-2})} = -8.6 - 0.76\log(S_{0.5-2})
\end{equation}

\noindent where $N(>S_{0.5-2})$ is in units of deg$^{-2}$ and $S_{0.5-2}$ is in units of ergs/cm$^2$/s. We note that this is in good agreement with other such fits to the X-ray number counts \citep[e.g.][]{2000Natur.404..459M,2001ApJ...551..624G}.

\begin{figure}
\centering
\includegraphics[width=\columnwidth]{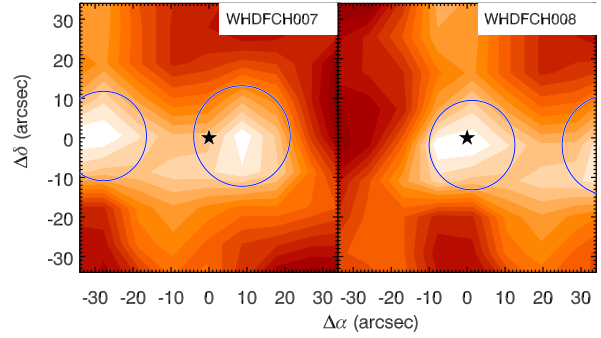}
\caption[]{Thumbnail images of the LABOCA 870\micron\ intensity map at the positions of the highly obscured ($N_H>10^{22}\mbox{ cm}^{-2}$) QSOs \ch{007} and \ch{008}. In each case the X-ray source position is marked by the star and nearby submm sources are marked by the blue circles (with the radii giving the estimated $2.5\Delta\theta$ positional accuracy). Both sources have closely associated submm emission.}
\label{f-lab0708}
\end{figure}

Based on the chosen $2.5\Delta\theta$ limits for identifying coincident sources, we find that two QSOs have possible LABOCA counterparts: \ch{007} (coincident with \lab{11}) and \ch{008} (coincident with \lab{05}). Fig.\ \ref{f-lab0708} shows thumbnail images of the LABOCA flux map at the positions of these X-ray sources, and $>3.2\sigma$ submm sources are seen close to the two QSOs. In each case, the star shows the location of the QSO and the blue circles show the $2.5\Delta\theta$ limit around nearby submm sources. The corrected Poisson probabilities that these are chance alignments are $P=0.029$ and $P=0.004$ (i.e. 2.9\% and 0.4\%) for \ch{007} and \ch{008} respectively. For comparison, we note that \citet{2007MNRAS.380..199I}, use a limit of $P\leq0.05$ as a constraint for secure alignments between submm and 24\micron\ sources.

These two alignments are with QSOs in our highly obscured sample, whilst we note that the other two highly obscured quasars, \ch{044} and -{\footnotesize099}, lie outside the imposed 2 mJy beam$^{-1}$ noise limit. It is interesting, therefore, that of the two highly obscured WHDF QSOs which lie in the central, low-noise area of the field, both appear to have bright submm counterparts.

\begin{figure}
\centering
\includegraphics[width=\columnwidth]{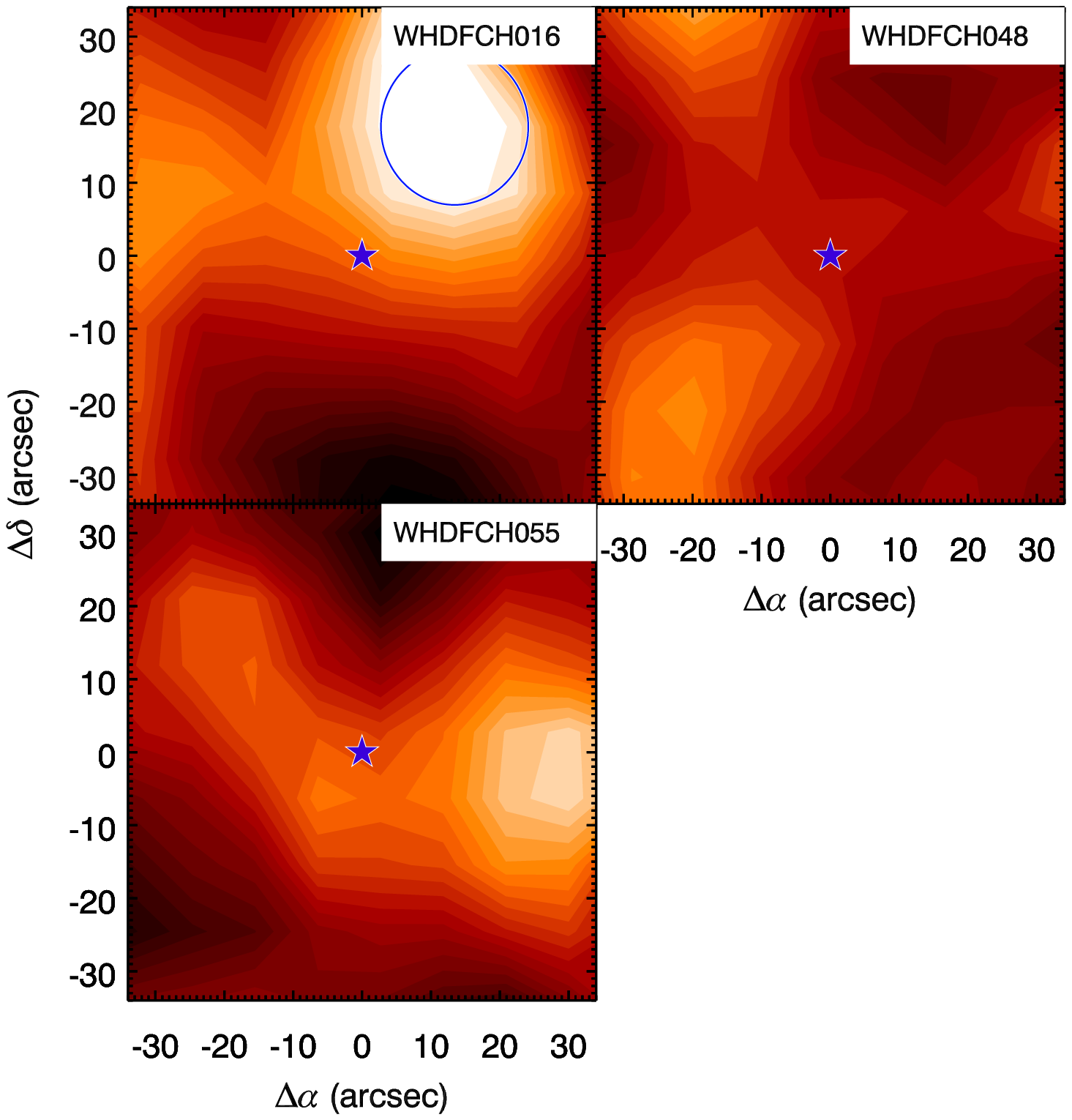}
\caption[LABOCA thumbnails of $N_{\mathrm{H}}\approx10^{22}$ cm\psq\ quasars]{As in Fig.\ \ref{f-lab0708}, but for sources characterised as mildly obscured with $N_{\mathrm{H}}\approx10^{22}$ cm\psq.}
\label{f-labmidabs}
\end{figure}

\begin{figure}
\centering
\includegraphics[width=\columnwidth]{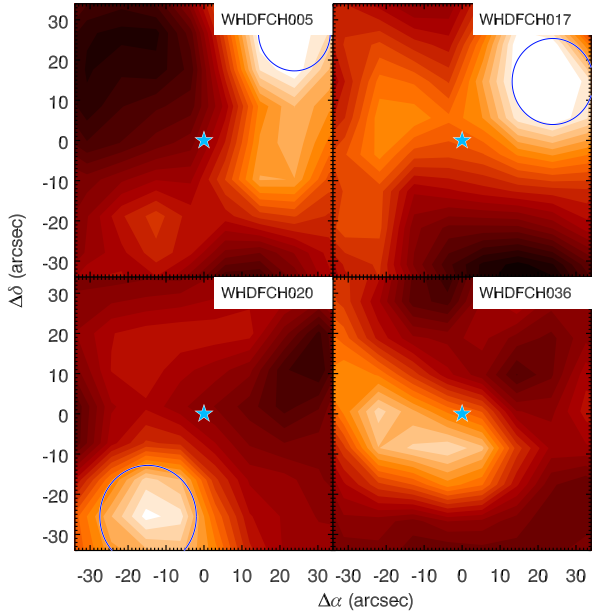}
\caption[LABOCA thumbnails of unobscured quasars]{As in Fig.\ \ref{f-lab0708}, but for sources characterised as unobscured with $N_{\mathrm{H}}<10^{22}$ cm\psq.}
\label{f-labunabs}
\end{figure}

In Figs.\ \ref{f-labmidabs} and \ref{f-labunabs}, we show thumbnail images for three $N_{\mathrm{H}}\approx10^{22}$ cm\psq\ QSOs and four unobscured QSOs, respectively. Again we note that the full samples have 6 and 8 sources respectively, but those QSOs not shown are rejected due to lying outside the $<$ 2 mJy beam$^{-1}$ noise region.

Some of the sources in these figures appear to lie close to bright areas, e.g.\  \ch{016} in Fig.\ \ref{f-labmidabs} or \ch{017} in Fig.\ \ref{f-labunabs}, however, none of these QSOs could be said to be \emph{coincident} with a peak in the map as was the case for \ch{007} and -{\footnotesize008}, lying as they do well outside the beam positional accuracy limits. We note that at the fluxes of the two submm sources associated with obscured AGN (i.e. $\approx3-4$ mJy/beam), our data is $\approx30\%$ complete, which could compromise the observation that the $N_H\lesssim10^{22}$ cm\psq\ QSOs are not associated with submm emission to the level of the $N_H\gtrsim10^{22}$ cm\psq\ QSOs. In the following section we therefore perform a stacking of each of the AGN populations to look for signatures of submm sources close to our detection threshold.

\subsubsection{Stacking}

To get an overall picture of the submm flux associated with our highly obscured, mildly obscured and unobscured quasar populations, we stack the submm flux maps for each of the populations. To ensure that the noisy fringe sources do not dominate the stacked flux there are two ways to proceed --- either to exclude the sources near the edge and  stack only the sources linearly within the $<2$ mJy/beam region (exploiting the relatively uniform noise across the centre), or to perform a noise-weighted stack which mitigates the effect of the noisy objects. Having performed both analyses, we find the results are not significantly affected by the choice of method. In the analysis presented here, for simplicity we use stacks incorporating only those sources within the $<2$ mJy/beam region, without any noise weighting.

\begin{figure}
\centering
\includegraphics[width=\columnwidth]{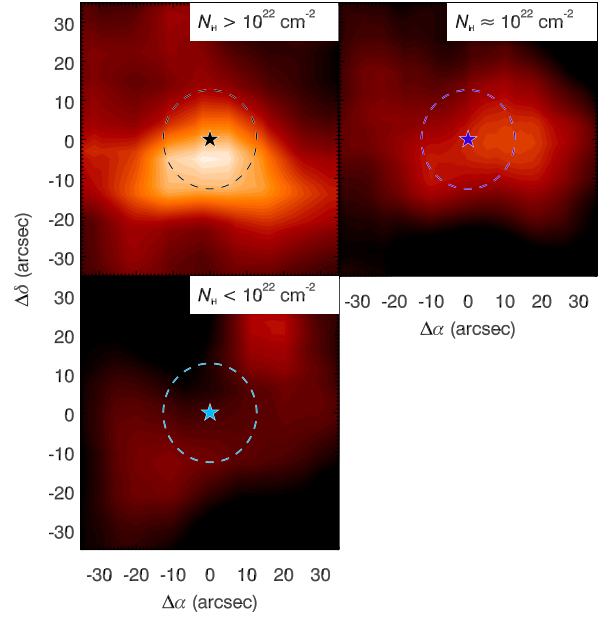}
\caption[Stacked LABOCA images of heavily obscured, mildly obscured and unobscured quasars]{Noise-weighted stacked intensity maps for the three populations with $N_{\mathrm{H}}>10^{22}$ cm\psq,  $N_{\mathrm{H}}\approx10^{22}$ cm\psq\ and $N_{\mathrm{H}}<10^{22}$ cm\psq. The dashed circle in each case shows the maximum search radius of $\Delta\theta=12''.7$ used in matching the QSOs and submm sources. Only the most highly obscured QSOs show significant stacked submm emission.}
\label{f-whdfstacks}
\end{figure}

The stacked flux images for the three QSO populations are shown in Fig.\ \ref{f-whdfstacks}, incorporating two sources in the $N_H>10^{22}$ cm\psq\ sample, three sources in the $N_H\approx10^{22}$ cm\psq\ sample and four sources in the $N_H<10^{22}$ cm\psq\ sample. Taking the peak value within our maximum search radius of $12.''7$, we find a peak signal of $S_{870\mu m} = 3.4\pm0.8$ mJy/beam at an angular distance of $5''$ from the QSO positions for the $N_H>10^{22}$ cm\psq\ stack. For the $N_H\approx10^{22}$ cm\psq\ and $N_H<10^{22}$ cm\psq\ stacks we find peak signals of $S_{870\mu m} = 1.9\pm0.8$ mJy/beam and $S_{870\mu m} = 0.7\pm0.5$ mJy/beam respectively (both being found at an angular distance of $\approx10''$ from the stack centre). These plots reinforce the observations based on individual X-ray sources, that the highly obscured quasars are the only sources to be associated with submm bright sources, without being affected by the completeness of the submm data at the $\approx3-4$ mJy/beam flux level. 

\subsubsection{Statistical power}

It is important to account for the statistical power afforded by the small numbers of objects involved in the current analysis. We therefore estimate the confidence limits of the sample size presented here (based on the tabulated small sample Poisson statistics provided by \citealt{1986ApJ...303..336G}) for the fraction of unobscured and obscured AGN associated with submm emission. Based on the sample of 2 obscured AGN, we estimate that our observations give a lower limit of 35\% on the percentage of such objects having associated $\gtrsim3$ mJy/beam submm emission (at the $1\sigma$ level). Conversely, the sample of $N_H<10^{22}$ cm\psq, gives an upper limit of 37\% on the percentage of these unobscured AGN being associated with $\gtrsim3$ mJy/beam submm emission (again at the $1\sigma$ level). Although the samples considered here are small and cannot fully constrain the relationship between AGN obscuration and submm emission, our analysis remains consistent with a model in which submm emission arises preferentially from more obscured AGN -- in agreement with the analyses of, for example, \citet{2004ApJ...611L..85P,2010ApJ...712.1287L,2011MNRAS.410..762H}.

\subsection{Future improvements}

Despite the relatively poor resolution of our submm data, it appears that each of the X-ray QSOs \ch{007} and \ch{008} has only a small chance of accidentally aligning with a submm source. It is notable that only the most obscured AGN show possible 870\micron\ detections and these include the Type 2 QSO, \ch{008}. Stacking the submm data as a function of X-ray absorption also provides further evidence that faint, X-ray obscured QSOs are preferentially submm bright.

Further tests of the reality of the associations between the two X-ray obscured QSOs and the LABOCA sources will soon be available from the 8.4GHz EVLA radio survey of the WHDF, which will have $\approx2\times$ improved spatial resolution compared to the LABOCA data. These data are still at the reduction stage and the results will be reported in a later paper\footnote{We note that since submission of this paper, both \ch{007} and \ch{008} have been detected in the 8.4GHz EVLA data.}.

In addition, we are seeking \emph{Herschel} observations at 250, 350 and 500\microns, which will constrain the SED and allow dust temperatures to be measured for the detected sources. If the unobscured QSOs are found to be detected in these shorter-wavelength bands, it could reveal whether any temperature difference exists between obscured and unobscured quasars, as was hypothesised by \citet{2011MNRAS.410..762H}.

Finally, we are also proposing to observe the two X-ray obscured QSOs with Atacama Large Millimetre/submillimetre Array (ALMA) in its extended configuration to  investigate further the submm counterparts of these sources. Not only will ALMA give better positional information but its $\approx0''.5$ angular  resolution will allow us to test if the associated submm sources are point-like or extended at $\la 3$kpc spatial resolution.

\section{Conclusions}

We have presented an 870\micron\ survey of the well-studied William Herschel Deep Field, reaching a depth of $<2$ mJy/beam over an area of 100 arcmin$^2$. In total, 11 sources have been detected at a significance of $\ge3.2\sigma$. From the noise distribution of the image, we estimate the number of false detections in the 100 arcmin$^2$ region to be $\approx0.2$, whilst the completeness in the same region is 50\% at a flux of 4.3 mJy/beam.

We find that 2 of the 11 submm sources are likely counterparts of X-ray-selected WHDF quasars: \ch{007}, an extremely hard QSO with $\Gamma\approx-1$ and \ch{008}, a type 2 QSO with only narrow optical emission lines, both consistent with being heavily obscured. We divide a sample of 15 WHDF quasars into three subsets, having $N_{\mathrm{H}}<10^{22}$, $N_{\mathrm{H}}\approx10^{22}$ and $N_{\mathrm{H}}>10^{22}$ cm\psq, and find that only the most obscured population shows any significant stacked 870\micron\ flux.

Although based on a relatively small sample, our findings are supportive of a model in which obscured AGN are submm bright and unobscured AGN are not. This picture of faint X-ray obscured AGN being preferentially stronger submm emitters is difficult to fit into the unified AGN model without some adaptation, given that both X-ray obscured and unobscured AGN would be expected to be equally strong submm emitters in this case. Nevertheless, the observational results of, for example,  \citet{2004ApJ...611L..85P,2005ApJ...632..736A, 2010ApJ...712.1287L,2011MNRAS.410..762H} and now the results presented here in the WHDF all tend to support the idea that obscured AGN are more frequently submm emitters than unobscured AGN.

From these results, two solutions have been proposed: either (a) AGN heating is a dominant component of dust heating, producing submm emission \citep[e.g.][]{1994MNRAS.268..235G,2002MNRAS.331..435W,2005MNRAS.359.1345G,2011MNRAS.410..762H} or (b) obscured AGN are an evolutionary concurrent phenomenon with strong star-formation and it is the star-formation that is the dominant mechanism for the dust heating in these objects \citep[e.g.][]{2004ApJ...611L..85P,2005Natur.434..738A,2009MNRAS.395.2189V,2010ApJ...712.1287L}.

Assuming the former, \citet{2011MNRAS.410..762H} argue that the contribution of obscured AGN to the submm background may be as much as $\approx40$\% (compared to $\approx13$\% in the unified/evolutionary model case). In addition, the same authors suggest that, should the primary driver of submm emission in these sources be heating from obscured AGN, then these may be the source of a significant fraction of the $\gtrsim10$ keV XRB, whilst the contribution would be much lower should heating from star-formation be the dominant mechanism. Other potential candidates still fall short of fully accounting for the XRB \citep[e.g.][]{2005MNRAS.357.1281W,2007A&A...463...79G,2009ApJ...696..110T} and in this context, quantifying the contribution from obscured AGN in the submm source population is an important goal. Although we are unable to inform fully on this based on the results here, these observations remain an important step in the process.

Upcoming radio data, as well as proposed observations in the submm and other bands, will enable further investigation of the submm properties of obscured and unobscured AGN in the WHDF. As discussed, Herschel observations will add important constraints to the SEDs of these objects and facilitate the measurement of dust temperatures, whilst the obscured QSOs will also make excellent targets for high resolution observations with ALMA. Crucially, such observations would not only improve on the accuracy of the submm source positions, but would also inform on whether the submm sources are extended, indicating a galaxy-wide starburst origin, or point-like and therefore more associated with the activity in the nucleus.

\section*{Acknowledgments}

We would like to thank J. Geach for assistance in reducing the submm data presented here and R. Hickox for comments on the text. MDH acknowledges the support of an STFC PhD Studentship grant, whilst RMB, NM and TS acknowledge the support of STFC funding. We also thank the anonymous referee for their contribution. This publication is based on data acquired with the Atacama Pathfinder Experiment (APEX). APEX is a collaboration between the Max-Planck-Institut fur Radioastronomie, the European Southern Observatory, and the Onsala Space Observatory.

\bibliographystyle{mnras_mod}
\bibliography{$HOME/Documents/lib/rmb}

\begin{thebibliography}{57}
\expandafter\ifx\csname natexlab\endcsname\relax\def\natexlab#1{#1}\fi

\bibitem[{{Alexander} {et~al.}(2011){Alexander}, {Bauer}, {Brandt}, {Daddi},
  {Hickox}, {Lehmer}, {Luo}, {Xue}, {Young}, {Comastri}, {Del Moro}, {Fabian},
  {Gilli}, {Goulding}, {Mainieri}, {Mullaney}, {Paolillo}, {Rafferty},
  {Schneider}, {Shemmer},,{Vignali}}]{2011ApJ...738...44A}
{Alexander} D.~M., {et~al.}, 2011, \apj, 738, 44

\bibitem[{{Alexander} {et~al.}(2003){Alexander}, {Bauer}, {Brandt},
  {Schneider}, {Hornschemeier}, {Vignali}, {Barger}, {Broos}, {Cowie},
  {Garmire}, {Townsley}, {Bautz}, {Chartas},,{Sargent}}]{2003AJ....126..539A}
{Alexander} D.~M., {et~al.}, 2003, \aj, 126, 539

\bibitem[{{Alexander} {et~al.}(2005{\natexlab{a}}){Alexander}, {Bauer},
  {Chapman}, {Smail}, {Blain}, {Brandt},,{Ivison}}]{2005ApJ...632..736A}
{Alexander} D.~M., {Bauer} F.~E., {Chapman} S.~C., {Smail} I., {Blain} A.~W.,
  {Brandt} W.~N., {Ivison} R.~J., 2005{\natexlab{a}}, \apj, 632, 736

\bibitem[{{Alexander} {et~al.}(2005{\natexlab{b}}){Alexander}, {Smail},
  {Bauer}, {Chapman}, {Blain}, {Brandt},,{Ivison}}]{2005Natur.434..738A}
{Alexander} D.~M., {Smail} I., {Bauer} F.~E., {Chapman} S.~C., {Blain} A.~W.,
  {Brandt} W.~N., {Ivison} R.~J., 2005{\natexlab{b}}, \nat, 434, 738

\bibitem[{{Almaini} {et~al.}(1999){Almaini},
  {Lawrence},,{Boyle}}]{1999MNRAS.305L..59A}
{Almaini} O., {Lawrence} A., {Boyle} B.~J., 1999, \mnras, 305, L59

\bibitem[{{Austermann} {et~al.}(2009){Austermann}, {Aretxaga}, {Hughes},
  {Kang}, {Kim}, {Lowenthal}, {Perera}, {Sanders}, {Scott}, {Scoville},
  {Wilson},,{Yun}}]{2009MNRAS.393.1573A}
{Austermann} J.~E., {et~al.}, 2009, \mnras, 393, 1573

\bibitem[{{Bielby} {et~al.}(2010){Bielby}, {Finoguenov}, {Tanaka}, {McCracken},
  {Daddi}, {Hudelot}, {Ilbert}, {Kneib}, {Le F{\`e}vre}, {Mellier}, {Nandra},
  {Petitjean}, {Srianand}, {Stalin},,{Willott}}]{2010A&A...523A..66B}
{Bielby} R.~M., {et~al.}, 2010, \aap, 523, A66+

\bibitem[{{B{\"o}hm} \&
  {Ziegler}(2007){B{\"o}hm},{Ziegler}}]{2007ApJ...668..846B}
{B{\"o}hm} A., {Ziegler} B.~L., 2007, \apj, 668, 846

\bibitem[{{Busswell} \&
  {Shanks}(2001){Busswell},{Shanks}}]{2001MNRAS.323...67B}
{Busswell} G.~S., {Shanks} T., 2001, \mnras, 323, 67

\bibitem[{{Carrera} {et~al.}(2011){Carrera}, {Page}, {Stevens}, {Ivison},
  {Dwelly}, {Ebrero},,{Falocco}}]{2011MNRAS.413.2791C}
{Carrera} F.~J., {Page} M.~J., {Stevens} J.~A., {Ivison} R.~J., {Dwelly} T.,
  {Ebrero} J., {Falocco} S., 2011, \mnras, 413, 2791

\bibitem[{{Chapman} {et~al.}(2009){Chapman}, {Blain}, {Ibata}, {Ivison},
  {Smail},,{Morrison}}]{2009ApJ...691..560C}
{Chapman} S.~C., {Blain} A., {Ibata} R., {Ivison} R.~J., {Smail} I., {Morrison}
  G., 2009, \apj, 691, 560

\bibitem[{{Comastri} {et~al.}(1995){Comastri}, {Setti},
  {Zamorani},,{Hasinger}}]{1995A&A...296....1C}
{Comastri} A., {Setti} G., {Zamorani} G., {Hasinger} G., 1995, \aap, 296, 1

\bibitem[{{Coppin} {et~al.}(2008{\natexlab{a}}){Coppin}, {Halpern}, {Scott},
  {Borys}, {Dunlop}, {Dunne}, {Ivison}, {Wagg}, {Aretxaga}, {Battistelli},
  {Benson}, {Blain}, {Chapman}, {Clements}, {Dye}, {Farrah}, {Hughes},
  {Jenness}, {van Kampen}, {Lacey}, {Mortier}, {Pope}, {Priddey}, {Serjeant},
  {Smail}, {Stevens},,{Vaccari}}]{2008MNRAS.384.1597C}
{Coppin} K., {et~al.}, 2008{\natexlab{a}}, \mnras,
  384, 1597

\bibitem[{{Coppin} {et~al.}(2009){Coppin}, {Smail}, {Alexander}, {Weiss},
  {Walter}, {Swinbank}, {Greve}, {Kovacs}, {De Breuck}, {Dickinson}, {Ibar},
  {Ivison}, {Reddy}, {Spinrad}, {Stern}, {Brandt}, {Chapman}, {Dannerbauer},
  {van Dokkum}, {Dunlop}, {Frayer}, {Gawiser}, {Geach}, {Huynh}, {Knudsen},
  {Koekemoer}, {Lehmer}, {Menten}, {Papovich}, {Rix}, {Schinnerer},
  {Wardlow},,{van der Werf}}]{2009MNRAS.395.1905C}
{Coppin} K.~E.~K., {et~al.}, 2009, \mnras, 395,
  1905

\bibitem[{{Coppin} {et~al.}(2008{\natexlab{b}}){Coppin}, {Swinbank}, {Neri},
  {Cox}, {Alexander}, {Smail}, {Page}, {Stevens}, {Knudsen}, {Ivison},
  {Beelen}, {Bertoldi},,{Omont}}]{2008MNRAS.389...45C}
{Coppin} K.~E.~K., {et~al.}, 2008{\natexlab{b}}, \mnras, 389, 45

\bibitem[{{Daddi} {et~al.}(2007){Daddi}, {Alexander}, {Dickinson}, {Gilli},
  {Renzini}, {Elbaz}, {Cimatti}, {Chary}, {Frayer}, {Bauer}, {Brandt},
  {Giavalisco}, {Grogin}, {Huynh}, {Kurk}, {Mignoli}, {Morrison},
  {Pope},,{Ravindranath}}]{2007ApJ...670..173D}
{Daddi} E., {et~al.}, 2007, \apj, 670, 173

\bibitem[{{Downes} {et~al.}(1986){Downes}, {Peacock},
  {Savage},,{Carrie}}]{1986MNRAS.218...31D}
{Downes} A.~J.~B., {Peacock} J.~A., {Savage} A., {Carrie} D.~R., 1986, \mnras,
  218, 31

\bibitem[{{Ellingson} {et~al.}(1991){Ellingson},
  {Yee},,{Green}}]{1991ApJ...371...49E}
{Ellingson} E., {Yee} H.~K.~C., {Green} R.~F., 1991, \apj, 371, 49

\bibitem[{{Fritz} {et~al.}(2009){Fritz},
  {B{\"o}hm},,{Ziegler}}]{2009MNRAS.393.1467F}
{Fritz} A., {B{\"o}hm} A., {Ziegler} B.~L., 2009, \mnras, 393, 1467

\bibitem[{{Gehrels}(1986)}]{1986ApJ...303..336G}
{Gehrels} N., 1986, \apj, 303, 336

\bibitem[{{Giacconi} {et~al.}(2001){Giacconi}, {Rosati}, {Tozzi}, {Nonino},
  {Hasinger}, {Norman}, {Bergeron}, {Borgani}, {Gilli},
  {Gilmozzi},,{Zheng}}]{2001ApJ...551..624G}
{Giacconi} R., {et~al.}, 2001, \apj, 551, 624

\bibitem[{{Gilli} {et~al.}(2007){Gilli},
  {Comastri},,{Hasinger}}]{2007A&A...463...79G}
{Gilli} R., {Comastri} A., {Hasinger} G., 2007, \aap, 463, 79

\bibitem[{{Granato} \& {Danese}(1994){Granato},{Danese}}]{1994MNRAS.268..235G}
{Granato} G.~L., {Danese} L., 1994, \mnras, 268, 235

\bibitem[{{Grimes} {et~al.}(2005){Grimes},
  {Rawlings},,{Willott}}]{2005MNRAS.359.1345G}
{Grimes} J.~A., {Rawlings} S., {Willott} C.~J., 2005, \mnras, 359, 1345

\bibitem[{{G{\"u}sten} {et~al.}(2006){G{\"u}sten}, {Nyman}, {Schilke},
  {Menten}, {Cesarsky},,{Booth}}]{2006A&A...454L..13G}
{G{\"u}sten} R., {Nyman} L.~{\AA}., {Schilke} P., {Menten} K., {Cesarsky} C.,
  {Booth} R., 2006, \aap, 454, L13

\bibitem[{{Hickox} \&
  {Markevitch}(2007){Hickox},{Markevitch}}]{2007ApJ...661L.117H}
{Hickox} R.~C., {Markevitch} M., 2007, \apjl, 661, L117

\bibitem[{{Hill} \&
  {Shanks}(2011{\natexlab{a}}){Hill},{Shanks}}]{2011MNRAS.414.1875H}
{Hill} M.~D., {Shanks} T., 2011{\natexlab{a}}, \mnras, 414, 1875

\bibitem[{{Hill} \&
  {Shanks}(2011{\natexlab{b}}){Hill},{Shanks}}]{2011MNRAS.410..762H}
{Hill} M.~D., {Shanks} T., 2011{\natexlab{b}}, \mnras, 410, 762

\bibitem[{{Ivison} {et~al.}(2007){Ivison}, {Greve}, {Dunlop}, {Peacock},
  {Egami}, {Smail}, {Ibar}, {van Kampen}, {Aretxaga}, {Babbedge}, {Biggs},
  {Blain}, {Chapman}, {Clements}, {Coppin}, {Farrah}, {Halpern}, {Hughes},
  {Jarvis}, {Jenness}, {Jones}, {Mortier}, {Oliver}, {Papovich},
  {P{\'e}rez-Gonz{\'a}lez}, {Pope}, {Rawlings}, {Rieke}, {Rowan-Robinson},
  {Savage}, {Scott}, {Seigar}, {Serjeant}, {Simpson}, {Stevens}, {Vaccari},
  {Wagg},,{Willott}}]{2007MNRAS.380..199I}
{Ivison} R.~J., {et~al.}, 2007, \mnras,
  380, 199

\bibitem[{{Krivonos} {et~al.}(2005){Krivonos}, {Vikhlinin}, {Churazov},
  {Lutovinov}, {Molkov},,{Sunyaev}}]{2005ApJ...625...89K}
{Krivonos} R., {Vikhlinin} A., {Churazov} E., {Lutovinov} A., {Molkov} S.,
  {Sunyaev} R., 2005, \apj, 625, 89

\bibitem[{{Lutz} {et~al.}(2010){Lutz}, {Mainieri}, {Rafferty}, {Shao},
  {Hasinger}, {Wei{\ss}}, {Walter}, {Smail}, {Alexander}, {Brandt}, {Chapman},
  {Coppin}, {F{\"o}rster Schreiber}, {Gawiser}, {Genzel}, {Greve}, {Ivison},
  {Koekemoer}, {Kurczynski}, {Menten}, {Nordon}, {Popesso}, {Schinnerer},
  {Silverman}, {Wardlow},,{Xue}}]{2010ApJ...712.1287L}
{Lutz} D., {et~al.}, 2010, \apj, 712, 1287

\bibitem[{{Mainieri} {et~al.}(2007){Mainieri}, {Hasinger}, {Cappelluti},
  {Brusa}, {Brunner}, {Civano}, {Comastri}, {Elvis}, {Finoguenov}, {Fiore},
  {Gilli}, {Lehmann}, {Silverman}, {Tasca}, {Vignali}, {Zamorani},
  {Schinnerer}, {Impey}, {Trump}, {Lilly}, {Maier}, {Griffiths}, {Miyaji},
  {Capak}, {Koekemoer}, {Scoville},
  {Shopbell},,{Taniguchi}}]{2007ApJS..172..368M}
{Mainieri} V., {et~al.}, 2007, \apjs, 172, 368

\bibitem[{{Mainieri} {et~al.}(2005){Mainieri}, {Rigopoulou}, {Lehmann},
  {Scott}, {Matute}, {Almaini}, {Tozzi},
  {Hasinger},,{Dunlop}}]{2005MNRAS.356.1571M}
{Mainieri} V., {et~al.}, 2005, \mnras, 356, 1571

\bibitem[{{Mart{\'{\i}}nez-Sansigre} {et~al.}(2009){Mart{\'{\i}}nez-Sansigre},
  {Karim}, {Schinnerer}, {Omont}, {Smith}, {Wu}, {Hill}, {Kl{\"o}ckner},
  {Lacy}, {Rawlings},,{Willott}}]{2009ApJ...706..184M}
{Mart{\'{\i}}nez-Sansigre} A., {et~al.}, 2009, \apj, 706, 184

\bibitem[{{Matsuda} {et~al.}(2011){Matsuda}, {Smail}, {Geach}, {Best},
  {Sobral}, {Tanaka}, {Nakata}, {Ohta}, {Kurk}, {Iwata}, {Bielby}, {Wardlow},
  {Bower}, {Ivison}, {Kodama}, {Yamada},
  {Mawatari},,{Casali}}]{2011MNRAS.416.2041M}
{Matsuda} Y., {et~al.}, 2011, \mnras, 416, 2041

\bibitem[{{McCracken} {et~al.}(2000{\natexlab{a}}){McCracken}, {Metcalfe},
  {Shanks}, {Campos}, {Gardner},,{Fong}}]{2000MNRAS.311..707M}
{McCracken} H.~J., {Metcalfe} N., {Shanks} T., {Campos} A., {Gardner} J.~P.,
  {Fong} R., 2000{\natexlab{a}}, \mnras, 311, 707

\bibitem[{{McCracken} {et~al.}(2000{\natexlab{b}}){McCracken}, {Shanks},
  {Metcalfe}, {Fong},,{Campos}}]{2000MNRAS.318..913M}
{McCracken} H.~J., {Shanks} T., {Metcalfe} N., {Fong} R., {Campos} A.,
  2000{\natexlab{b}}, \mnras, 318, 913

\bibitem[{{Metcalfe} {et~al.}(1995){Metcalfe}, {Fong},,{Shanks}}]{metcalfe95}
{Metcalfe} N., {Fong} R., {Shanks} T., 1995, \mnras, 274, 769

\bibitem[{{Metcalfe} {et~al.}(2001){Metcalfe}, {Shanks}, {Campos},
  {McCracken},,{Fong}}]{metcalfe01}
{Metcalfe} N., {Shanks} T., {Campos} A., {McCracken} H.~J., {Fong} R., 2001,
  \mnras, 323, 795

\bibitem[{{Metcalfe} {et~al.}(2006){Metcalfe}, {Shanks}, {Weilbacher},
  {McCracken}, {Fong},,{Thompson}}]{metcalfe06}
{Metcalfe} N., {Shanks} T., {Weilbacher} P.~M., {McCracken} H.~J., {Fong} R.,
  {Thompson} D., 2006, \mnras, 370, 1257

\bibitem[{{Mushotzky} {et~al.}(2000){Mushotzky}, {Cowie},
  {Barger},,{Arnaud}}]{2000Natur.404..459M}
{Mushotzky} R.~F., {Cowie} L.~L., {Barger} A.~J., {Arnaud} K.~A., 2000, \nat,
  404, 459

\bibitem[{{Page} {et~al.}(2004){Page}, {Stevens},
  {Ivison},,{Carrera}}]{2004ApJ...611L..85P}
{Page} M.~J., {Stevens} J.~A., {Ivison} R.~J., {Carrera} F.~J., 2004, \apjl,
  611, L85

\bibitem[{{Pope} {et~al.}(2006){Pope}, {Scott}, {Dickinson}, {Chary},
  {Morrison}, {Borys}, {Sajina}, {Alexander}, {Daddi}, {Frayer},
  {MacDonald},,{Stern}}]{2006MNRAS.370.1185P}
{Pope} A., {et~al.}, 2006, \mnras, 370, 1185

\bibitem[{{Schuller} {et~al.}(2010){Schuller}, {Nord}, {Vlahakis}, {Albrecht},
  {Beelen}, {Bertoldi}, {Mueller},,{Schaaf}}]{2010BoA}
{Schuller} F., {Nord} M., {Vlahakis} C., {Albrecht} M., {Beelen} A., {Bertoldi}
  F., {Mueller} S., {Schaaf} R., 2010, {BoA User Manual}

\bibitem[{{Shanks} {et~al.}(1991){Shanks}, {Georgantopoulos}, {Stewart},
  {Pounds}, {Boyle},,{Griffiths}}]{1991Natur.353..315S}
{Shanks} T., {Georgantopoulos} I., {Stewart} G.~C., {Pounds} K.~A., {Boyle}
  B.~J., {Griffiths} R.~E., 1991, \nat, 353, 315

\bibitem[{{Siemiginowska} {et~al.}(2010){Siemiginowska}, {Burke}, {Aldcroft},
  {Worrall}, {Allen}, {Bechtold}, {Clarke},,{Cheung}}]{2010ApJ...722..102S}
{Siemiginowska} A., {Burke} D.~J., {Aldcroft} T.~L., {Worrall} D.~M., {Allen}
  S., {Bechtold} J., {Clarke} T., {Cheung} C.~C., 2010, \apj, 722, 102

\bibitem[{{Siringo} {et~al.}(2009){Siringo}, {Kreysa}, {Kov{\'a}cs},
  {Schuller}, {Wei{\ss}}, {Esch}, {Gem{\"u}nd}, {Jethava}, {Lundershausen},
  {Colin}, {G{\"u}sten}, {Menten}, {Beelen}, {Bertoldi},
  {Beeman},,{Haller}}]{2009A&A...497..945S}
{Siringo} G., {et~al.}, 2009, \aap, 497, 945

\bibitem[{{Smail} {et~al.}(2004){Smail}, {Chapman},
  {Blain},,{Ivison}}]{2004ApJ...616...71S}
{Smail} I., {Chapman} S.~C., {Blain} A.~W., {Ivison} R.~J., 2004, \apj, 616,
  71

\bibitem[{{Smith} \& {Heckman}(1990){Smith},{Heckman}}]{1990ApJ...348...38S}
{Smith} E.~P., {Heckman} T.~M., 1990, \apj, 348, 38

\bibitem[{{Treister} {et~al.}(2009){Treister},
  {Urry},,{Virani}}]{2009ApJ...696..110T}
{Treister} E., {Urry} C.~M., {Virani} S., 2009, \apj, 696, 110

\bibitem[{{Vallb\'{e}-Mumbru}(2004)}]{2004PhDT.VallbeMumbru}
{Vallb\'{e}-Mumbru} M., 2004, PhD thesis, Durham University

\bibitem[{{Valtchanov} {et~al.}(2011){Valtchanov}, {Virdee}, {Ivison},
  {Swinyard}, {van der Werf}, {Rigopoulou}, {da Cunha}, {Lupu}, {Benford},
  {Riechers}, {Smail}, {Jarvis}, {Pearson}, {Gomez}, {Hopwood}, {Altieri},
  {Birkinshaw}, {Coia}, {Conversi}, {Cooray}, {de Zotti}, {Dunne}, {Frayer},
  {Leeuw}, {Marston}, {Negrello}, {Portal}, {Scott}, {Thompson}, {Vaccari},
  {Baes}, {Clements}, {Micha{\l}owski}, {Dannerbauer}, {Serjeant}, {Auld},
  {Buttiglione}, {Cava}, {Dariush}, {Dye}, {Eales}, {Fritz}, {Ibar}, {Maddox},
  {Pascale}, {Pohlen}, {Rigby}, {Rodighiero}, {Smith}, {Temi}, {Carpenter},
  {Bolatto}, {Gurwell},,{Vieira}}]{2011MNRAS.415.3473V}
{Valtchanov} I., {et~al.},
  2011, \mnras, 415, 3473

\bibitem[{{Vignali} {et~al.}(2009){Vignali}, {Pozzi}, {Fritz}, {Comastri},
  {Gruppioni}, {Bellocchi}, {Fiore}, {Brusa}, {Maiolino}, {Mignoli}, {La
  Franca}, {Pozzetti}, {Zamorani},,{Merloni}}]{2009MNRAS.395.2189V}
{Vignali} C., {et~al.}, 2009, \mnras, 395, 2189

\bibitem[{{Wei{\ss}} {et~al.}(2009){Wei{\ss}}, {Kov{\'a}cs}, {Coppin}, {Greve},
  {Walter}, {Smail}, {Dunlop}, {Knudsen}, {Alexander}, {Bertoldi}, {Brandt},
  {Chapman}, {Cox}, {Dannerbauer}, {De Breuck}, {Gawiser}, {Ivison}, {Lutz},
  {Menten}, {Koekemoer}, {Kreysa}, {Kurczynski}, {Rix}, {Schinnerer},,{van der
  Werf}}]{2009ApJ...707.1201W}
{Wei{\ss}} A., {et~al.}, 2009, \apj, 707, 1201

\bibitem[{{Willott} {et~al.}(2002){Willott}, {Rawlings},
  {Archibald},,{Dunlop}}]{2002MNRAS.331..435W}
{Willott} C.~J., {Rawlings} S., {Archibald} E.~N., {Dunlop} J.~S., 2002,
  \mnras, 331, 435

\bibitem[{{Worsley} {et~al.}(2006){Worsley}, {Fabian}, {Bauer}, {Alexander},
  {Brandt},,{Lehmer}}]{2006MNRAS.368.1735W}
{Worsley} M.~A., {Fabian} A.~C., {Bauer} F.~E., {Alexander} D.~M., {Brandt}
  W.~N., {Lehmer} B.~D., 2006, \mnras, 368, 1735

\bibitem[{{Worsley} {et~al.}(2005){Worsley}, {Fabian}, {Bauer}, {Alexander},
  {Hasinger}, {Mateos}, {Brunner}, {Brandt},,{Schneider}}]{2005MNRAS.357.1281W}
{Worsley} M.~A., {et~al.}, 2005, \mnras, 357, 1281

\end{thebibliography}

\label{lastpage}

\end{document}